\pdfoutput=1
\documentclass[preprint,12pt]{elsarticle}
\usepackage{amsmath}
\usepackage{url}
\usepackage{graphicx}
\usepackage{bm}
\usepackage{here}
\usepackage{caption}
\usepackage[subrefformat=parens]{subcaption}
\newcounter{bla}

\begin{document}
\renewcommand{\include}[1]{}
\renewcommand\documentclass[2][]{}
\setcounter{tocdepth}{2}
\thispagestyle{empty}
\journal{Computer Physics Communications}

\begin{frontmatter}



\title{Code O-SUKI: Simulation of Direct-Drive Fuel Target Implosion in Heavy Ion Inertial Fusion}


\author[a]{R. Sato}
\author[a]{S. Kawata\corref{author}}
\author[a]{T. Karino}
\author[a]{K. Uchibori}
\author[a]{T. Iinuma}
\author[a]{H. Katoh}
\author[b]{A. I. Ogoyski}

\cortext[author] {Corresponding author.\\\textit{E-mail address:} kwt@cc.utsunomiya-u.ac.jp}
\address[a]{Graduate School of Engineering, Utsunomiya University, Utsunomiya 321-8585, Japan}
\address[b]{Department of Physics, Varna Technical University, Varna 9010, Bulgaria}

\begin{abstract}

The Code O-SUKI is an integrated 2-dimensional (2D) simulation program system for a fuel implosion, ignition and burning of a direct-drive nuclear-fusion pellet in heavy ion beam (HIB) inertial confinement fusion (HIF). The Code O-SUKI consists of the four programs of the HIB illumination and energy deposition program of OK3 (Comput. Phys. Commun. 181, 1332 (2010)), a Lagrangian fluid implosion program, a data conversion program, and an Euler fluid implosion, ignition and burning program. The OK3 computes the multi-HIBs irradiation onto a spherical fuel target. One HIB is divided into many beamlets in OK3. Each heavy ion beamlet deposits its energy along the trajectory in a deposition layer depending on the particle energy. The OK3 also has a function of a wobbling motion of the HIB axis oscillation, and the HIBs energy deposition spatial detail profile is obtained inside the energy absorber of the fuel target. The spherical target implosion 2D behavior is computed by the 2D Lagrangian fluid code coupled with OK3, until just before the void closure time of the fuel implosion. After that, all the data by the Lagrangian implosion code are converted to them for the Eulerian code. The fusion Deuterium (D)-Tritium (T) fuel and the inward moving heavy tamping material are imploded and deformed seriously at the stagnation phase. The Euler fluid code is appropriate to simulate the fusion fuel compression, ignition and burning.  The Code O-SUKI 2D simulation system provides a capability to compute and to study the HIF target implosion dynamics. 

\end{abstract}


\end{frontmatter}



\noindent
{\bf Program summary}

\begin{small}
\noindent
{\em Program Title:}  O-SUKI                                        \\
{\em Licensing provisions: CC BY NC 3.0 }                                   \\
{\em Programming language:} C++                                 \\
{\em Computer:} PC(Pentium 4, 1 GHz or more recommended)\\
{\em RAM:} 3072 MBytes\\
{\em Operating system:} UNIX\\
{\em Journal reference of previous version: No}                  \\
{\em Nature of problem:}     
The nuclear fusion energy would provide one of energy resources for our human society. In this paper we focus on heavy ion beam (HIB) inertial confinement fusion (HIF). A spherical deuterium (D) - tritium (T) fuel pellet, whose radius may be about several mm, is irradiated by HIBs to be compressed to about a thousand times of the solid density. The DT fuel temperature reaches $\sim$5-10KeV to be ignited to release the DT fusion energy. The typical HIBs total input energy is several MJ, and the HIBs pulse length is about a few tens of ns. The DT fuel compression uniformity is essentially important to release the sufficient fusion energy output. The DT fuel pellet implosion non-uniformity should be kept less than a few \%.  The O-SUKI code system provides an integrated tool to simulate the HIF DT fuel pellet implosion, ignition and burning. The HIBs energy deposition detail profile is computed by the OK3 code (Comput. Phys. Commun. 181, 1332 (2010)) in an energy absorber outer layer, which covers the DT fuel spherical shell. The DT fuel is compressed to the high density, and so the DT fuel spatial deformation may be serious at the DT fuel stagnation. Therefore, the O-SUKI system employs a Lagrangian fluid code first to simulate the DT fuel implosion phase until just before the stagnation. Then all the simulation data from the Lagrangian code are converted to them for the Euler fluid code, in which the DT fuel ignition and burning are simulated. \\
{\em Solution method:}     
In the two fluids codes (Lagrangian and Euler fluid codes) in the O-SUKI system the three-temperature fluid model (J. Appl. Phys. 60, 898 (1986)) is employed to simulate the pellet dynamics in HIF. The HIBs energy deposition detail profile is computed by the OK3 code (Comput. Phys. Commun. 181, 1332 (2010)). 
\\
{\em Additional comments including Restrictions and Unusual features:} No\\
   \\

\end{small}

\section{Introduction}\label{sec:1}

In inertial confinement fusion (ICF) a deuterium (D) - tritium (T) fuel target implosion, ignition and burning are essentially important to release a sufficient fusion energy output. In the nuclear fusion two nuclei of D and T are fused once. The DT fusion reaction creates He and neutron, and releases the energy of the mass defect as their kinetic energy in the DT reaction. In ICF a few mg DT in a fuel pellet is first heated up to $\sim$5-10KeV by an input driver energy, for example, lasers or heavy ion beams (HIBs) or pulse power \cite{ICFBook}. Especially, the solid DT fuel density must be compressed to about a thousand times of the solid density to reduce the input energy and also to realize controlled fusion reactions. In addition, the ion temperature of the compressed DT must reach $\sim$5-10 KeV. In order to compress the DT fuel stably to the high density, the implosion non-uniformity should be less than a few percent\cite {kwtANDniu} The central issues of the fuel implosion in ICF includes how to realize the uniform implosion, how we can control the driver beam energy deposition to compress the DT fuel to the high density, and consequently how to keep the implosion stable during the fuel implosion. The O-SUKI code system provides an integrated computer simulation tool to study the DT fuel implosion, ignition and burning in heavy ion inertial confinement fusion (HIF). 
 
 The heavy ion beam (HIB) fusion has been proposed in 1970s. The recent HIF activities and reviews are found in Refs. \cite{Kawata1, IGHoffman}. The HIF reactor designs were also proposed\cite {bohne, ymk, moir}. HIB ions deposit their energy inside of materials, and the interaction of the HIB ions with the materials are well understood \cite {ziegler, mehlhorn}. The HIB ion interaction with a material is explained and defined well by the classical Coulomb collision and a plasma wave excitation in the material plasma. The HIB ions deposit all the HIB ion energy inside of the material. The HIB energy deposition length is typically the order of $\sim$mm in an HIF fuel target depending on the HIB ion energy and the material. When several MJ of the HIB energy is deposited in the material in an inertial confinement fusion (ICF) fuel target, the temperature of the energy deposition layer plasma becomes about 300 eV or so. The peak temperature or the peak plasma pressure appears near the HIB ion stopping area by the Bragg peak effect, which comes from the nature of the Coulomb collision. The total stopping range would be normally wide and the order of $\sim$mm inside of the material. An indirect drive target was also proposed in Ref. \cite {CM}. 
 
          In ICF, a driver efficiency and its repetitive operation with several Hz $\sim$ 20 Hz or so are essentially important to constitute an ICF reactor system. HIB driver accelerators have a high driver energy efficiency of $\sim$30-40 \% from the electricity to the HIB energy. In general, high-energy accelerators have been operated repetitively daily. The high driver efficiency relaxes the requirement for the fuel target gain. In HIF the target gain of 30$\sim$50 allows us to construct a HIF fusion reactor system, and 1MkW of the electricity output would be realized with the repetition rate of $\sim$10$\sim$15 Hz. 
          
          The HIB accelerator also has a high controllability to define the ion energy, the HIB pulse shape, the HIB pulse length and the HIB number density or current as well as the beam axis. The HIB axis could be also controlled or oscillated with a high frequency\cite {arnold, piriz, qin}. The controlled wobbling motion of the HIB axis is one of remarkable preferable points in HIF, and would contribute to smooth the HIBs illumination non-uniformity on a DT fuel target and to mitigate the Rayleigh-Taylor (R-T) instability growth in the HIF fuel target implosion\cite {KSrad, KSdynamic, KKrobust}. In the OK3 code the HIBs wobbling capability is also installed to study the wobbling ion beam energy deposition. 
        
          The relatively large density gradient scale length is created in the HIBs energy deposition region in an DT fuel target, and it also contribute to reduce the R-T instability growth rate especially for shorter wavelength modes\cite {bodner, takabe}. So in the HIF target implosion longer wavelength modes should be focused for the target implosion uniformity. 

     In general the target implosion non-uniformity is introduced by a driver beams' illumination non-uniformity, an imperfect target sphericity, a non-uniform target density, a target alignment error in a fusion reactor, et al. The target implosion should be also robust against the implosion non-uniformities for the stable reactor operation. 

         In the HIBs energy deposition region in a DT fuel target a wide density valley appears, and in the density valley a part of the HIBs deposited energy is converted to the radiation and the radiation is confined in the density valley \cite{Sasaki}. The converted and confined radiation energy is not negligible, and it would be the order of $\sim$100 kJ in a HIF reactor-size DT target. The confined radiation in the density valley contributes also to reduce the non-uniformity of the HIBs energy deposition. 
         
 The HIB uniform illumination was also studied, and the target implosion uniformity requirement requests the minimum HIB number: details HIBs energy deposition on a direct-drive DT fuel target shows that the minimum HIBs number would be the 32 beams \cite {Miyazawa}. The detail HIBs illumination on a HIF DT target is computed by a computer code of OK3 \cite{ogoyski1, ogoyski2, ogoyski3}. The HIBs illumination non-uniformity is also studied in detail. One of the study results shows that a target misalignment of $\sim$100$\mu$m is tolerable in fusion reactor to release the HIF energy stably.
 
 The DT fuel implosion is simulated until just before the void closure time by the Lagrangian code, which couples with the OK3 code to include the time-dependent HIBs energy deposition profile in the target energy absorber layer. The Lagrange code data are converted to the data imported to the Euler code, which is robust against the target fuel and material deformation. The DT fuel ignition and burning are simulated further by the Euler fluid code. The O-SUKI code system simulates the 2D HIF target implosion dynamics, and would contribute to release the fusion energy stably and in a robust way for our human society.



\section{O-SUKI code algorithm description}
\par

\subsection{O-SUKI code structure}
     The O-SUKI code system is an integrated DT fuel implosion code in HIF, and consists of four parts: The HIBs illumination code of OK3 \cite{ogoyski3}, the Lagrangian fluid code \cite{Schulz}, the data conversion code from the Lagrangian code to the Euler code, and Euler code. The fluid model is the three-temperature model in Ref. \cite{Tahir}. The detail information on OK3 is presented in Refs. \cite{ogoyski1, ogoyski2, ogoyski3}. The Lagrangian fluid code, the data conversion code and the Euler code are described below in detail. 
     
     In the Lagrangian fluid code the spatial meshes move together with the fluid motion \cite{Schulz}. However the mass and energy conservations are well described, the Lagrange meshes can not follow the fluid large deformation. On the other hand, the Euler meshes are fixed to the space, and the fluid moves through the meshes. Therefore, just before the void closure time, that is, the stagnation phase, the Lagrangian code is used to simulate the DT fuel implosion. After the void closure time, the Euler code is employed to simulate the DT fuel further compression, ignition and burning. Between the Lagrangian code and the Euler code the data should be converted by the data conversion code. 

	All the simulation process is performed in its integrated way by using the script of "O-SUKIcode\_start.sh". The processes executed by this shell script are as follows.\\
1. Make the stack size infinite.\\
2. Change the permission of shell scripts to executable. \\
3. Compile the main function of the Lagrangian code and execute it.\\
4. If any problems do not appear during the calculation of the Lagrangian code, compile the main function of the data conversion code and execute it.\\
5. If there is no problem during the data conversion, compile the main function of the Euler code and execute it.\\

\subsection{Steps in Lagrangian code}\par
     The Lagrangian code has the following steps: 

\begin{enumerate}
\item Initialize the variables. \par
\item Calculation of time step size.\par
\item Calculation of coordinates.\par
\item Solve equation of motion. \par
\item Solve density by equation of continuity.\par
\item Calculation of artificial viscosity.\par
\item Transfer the data to the OK3. \par
\item Calculation of energy deposition distribution in code OK3. For details of the OK3, see the ref.\cite{ogoyski1,ogoyski2,ogoyski3}. \par
\item Solve energy equations\par
\item Calculation of heat conduction\par
\item Calculation of temperature relaxation among three temperatures.\par
\item Solve equation of state\par
\item Save the results.\par
\item End the Lagrangian calculation right before the void closure.\par
\item Transfer the data to converting code. \par
\end {enumerate}

\subsection{Data Conversion code from Lagrangian fluid code to Euler fluid code}

\begin {enumerate}
\item Read variables saved in Lagrangian code.\par
\item Generate the Eulerian mesh.\par
\item Calculate the interpolation of the physical quantity to them on the Eulerian mesh.\par
\item Write the converted data to the Eulerian code.\par
\end {enumerate}

\subsection{Steps in Eulerian code}

\begin {enumerate}
\item Read the mesh number from the converted data and define the each matrices.\par
\item Initialize the variables.\par
\item Calculation of time step size.\par
\item Solve equation of motion. \par
\item Track the material boundaries of DT, Al and Pb.\par
\item Linearly interpolate the boundary lines and transcribe them on the Eulerian code. \par
\item Discriminate the materials by using the transferred boundary line. \par
\item Solve density by equation of continuity.\par
\item Calculate artificial viscosity.\par
\item Solve energy equations\par
\item Calculation of fusion reaction.\par
\item Calculation of heat conduction\par
\item Calculation of temperature relaxation among three temperatures.\par
\item Solve equation of state.\par
\item Save the results.\par
\item End.
\end{enumerate}
\documentclass[preprint,12pt]{elsarticle}

\section{Included files}

The coordinates in the Lagrangian fluid simulation code are as shown below (see Fig. \ref{coordinate_fluid})．The discretization method in Ref. \cite{Schulz} is employed in the Lagrangian fluid code. 
\begin{eqnarray*}
	R=R(k,l,t)\\
	Z=Z(k,l,t)
\end{eqnarray*}
The position vector $ {\bm R} $ and the vector $ {\bar {\bm R}} $ are introduced as follows.
\begin{equation*}
	{\bm R}=
	\begin{bmatrix}
	R\\
	Z
	\end{bmatrix}
	\ ,\ {\bar {\bm R}}=
	\begin{bmatrix}
	Z\\
	-R
	\end{bmatrix}
\end{equation*}

\begin{figure}[H]
	\centering
	\includegraphics[height=8cm]{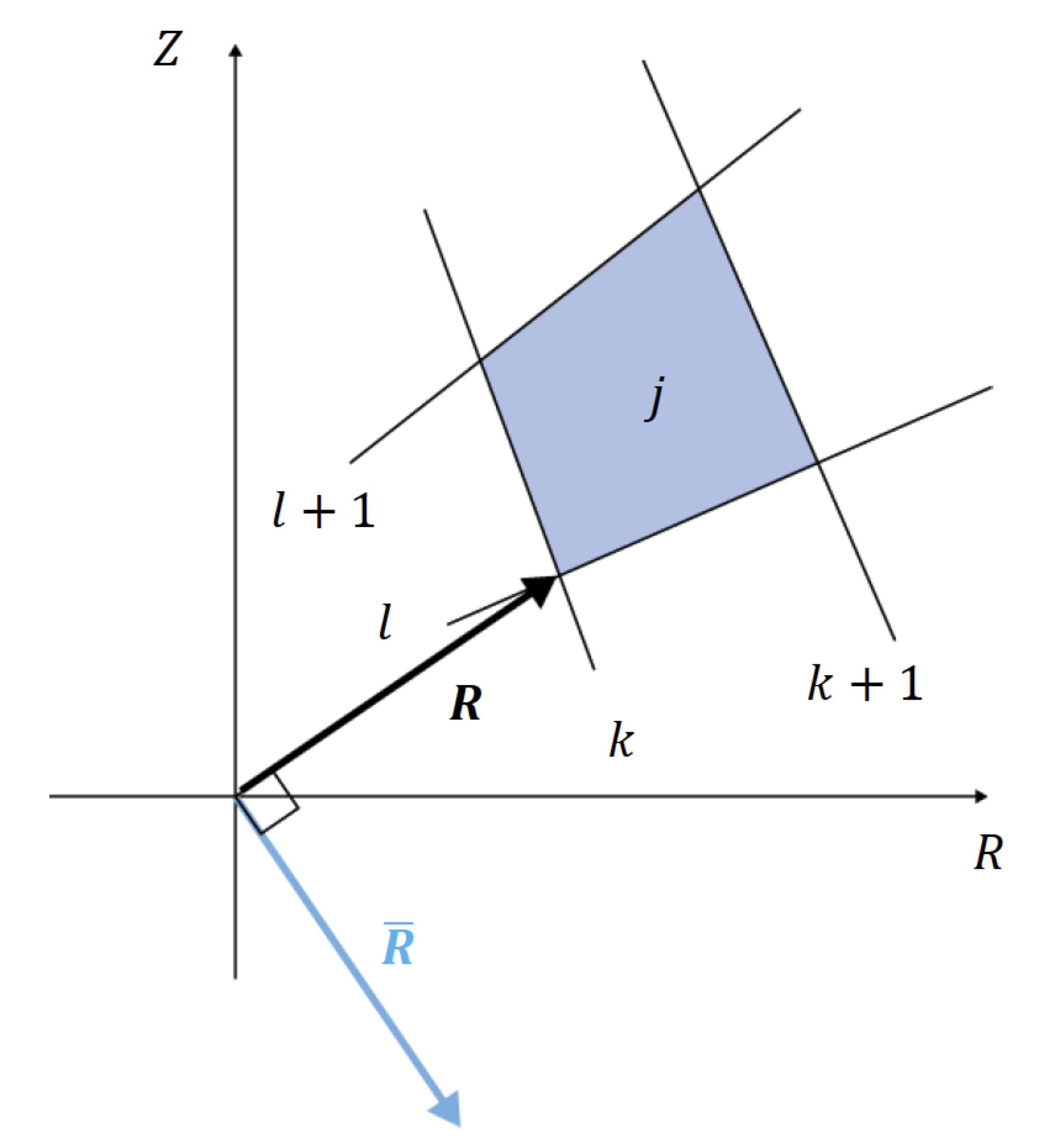}
	\caption{Lagrangian coordinate.}\label{coordinate_fluid}
\end{figure}

The definition points of the discretized physical quantities in the Lagrange and Euler codes are presented in Figs. \ref {definition_La} and \ref{definition_Eu}, respectively. The subscripts $ k $ and $ l $ correspond to the positions in space, and the subscript $ n $ corresponds to time $n\times dt$. The displacement amounts in the $k$ and $l$ directions are defined as follows.
	\begin{eqnarray}
		dR^n_{k+\frac{1}{2},l}=R^n_{k+1,l}-R^n_{k,l}\\
		dZ^n_{k,l+\frac{1}{2}}=Z^n_{k,l+1}-Z^n_{k,l}
	\end{eqnarray}	

	\begin{eqnarray*}
		&&\begin{cases}
			\Delta R^n_{k+\frac{1}{2},l}=R^n_{k+1,l}-R^n_{k,l}\\
			\delta R^n_{k,l+\frac{1}{2}}=R^n_{k,l+1}-R^n_{k,l}
		\end{cases}\\
		&&\begin{cases}
			\Delta Z^n_{k+\frac{1}{2},l}=Z^n_{k+1,l}-Z^n_{k,l}\\
			\delta Z^n_{k,l+\frac{1}{2}}=Z^n_{k,l+1}-Z^n_{k,l}
		\end{cases}\\
		&&\begin{cases}
			\Delta {\bm R}^n_{k+\frac{1}{2},l}={\bm R}^n_{k+1,l}-{\bm R}^n_{k,l}\\
			\delta {\bm R}^n_{k,l+\frac{1}{2}}={\bm R}^n_{k,l+1}-{\bm R}^n_{k,l}
		\end{cases}
	\end{eqnarray*}

\begin{figure}[H]
	\centering
	\includegraphics[height=8cm]{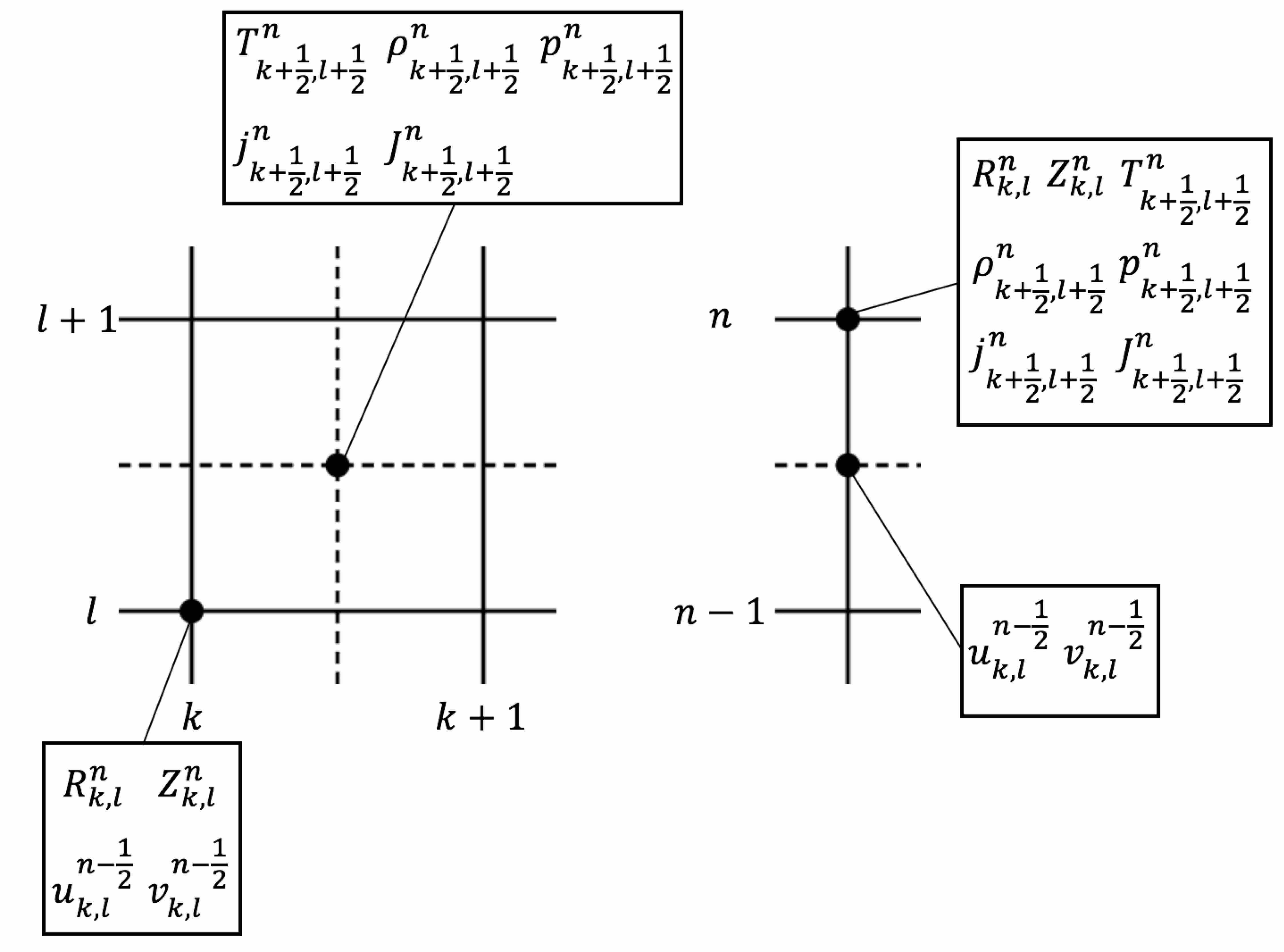}
	\caption{Definition points of discretized physical quantities in the Lagrangian code.}\label{definition_La}
	\end{figure}
	
\begin{figure}[H]
	\centering
	\includegraphics[height=8cm]{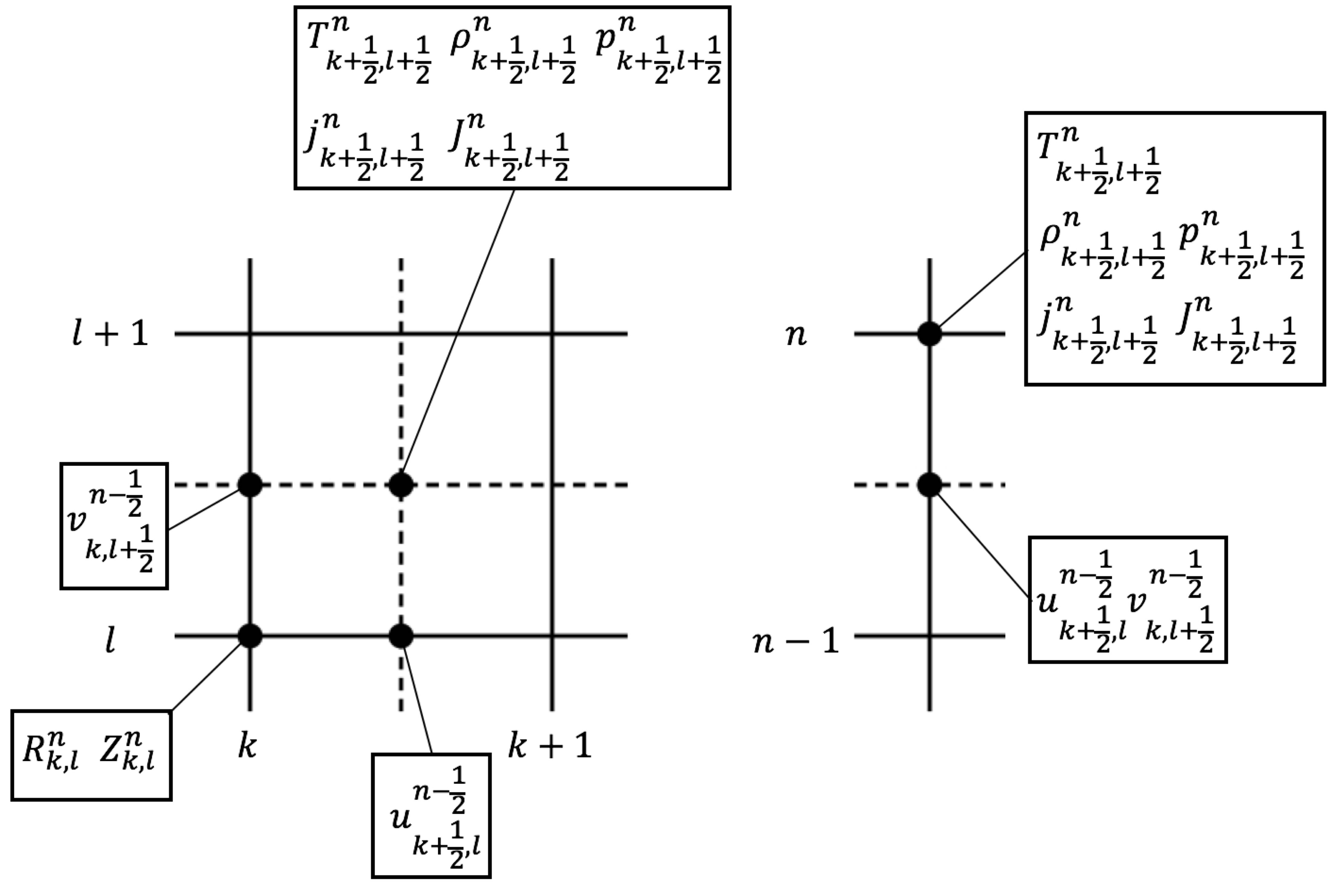}
	\caption{Definition points of discretized physical quantities in the Eulerian code.}\label{definition_Eu}
	\end{figure}
\end{document}